\documentclass[prd,twocolumn,amssymb,showpacs,nofootinbib,superscriptaddress,aps]{revtex4}
\usepackage{graphicx}
\usepackage{bm}
\usepackage{amsmath}
\usepackage{amssymb}
\usepackage{hyperref}

\begin{document}
\title{Visualization of semileptonic form factors from lattice QCD}

\author{C.~Bernard}
\affiliation{Department of Physics, Washington University, 
St.~Louis, Missouri, USA}
\author{C.~DeTar}
\affiliation{Physics Department, University of Utah, Salt Lake City, Utah, USA}
\author{M.~Di Pierro}
\affiliation{School of Computer Science, Telecommunications and 
Information Systems, DePaul University, Chicago, Illinois, USA}
\author{A.X.~El-Khadra}
\affiliation{Physics Department, University of Illinois, 
Urbana, Illinois, USA}
\author{R.T.~Evans}
\affiliation{Physics Department, University of Illinois, 
Urbana, Illinois, USA}
\author{E.D.~Freeland}
\affiliation{Liberal Arts Department, The School of the Art Institute of 
Chicago, Chicago, Illinois, USA }
\author{E.~Gamiz}
\affiliation{Physics Department, University of Illinois, 
Urbana, Illinois, USA}
\author{Steven~Gottlieb}
\affiliation{Department of Physics, Indiana University, 
Bloomington, Indiana, USA}
\author{U.M.~Heller}
\affiliation{American Physical Society, Ridge, New York, USA}
\author{J.E.~Hetrick}
\affiliation{Physics Department, University of the Pacific, 
Stockton, California, USA}
\author{A.S.~Kronfeld}
\affiliation{Fermi National Accelerator Laboratory, Batavia, Illinois, USA}
\author{J.~Laiho}
\affiliation{Department of Physics, Washington University, 
St.~Louis, Missouri, USA}
\author{L.~Levkova}
\affiliation{Physics Department, University of Utah, Salt Lake City, Utah, USA}
\author{P.B.~Mackenzie}
\affiliation{Fermi National Accelerator Laboratory, Batavia, Illinois, USA}
\author{M.~Okamoto}
\affiliation{Fermi National Accelerator Laboratory, Batavia, Illinois, USA}
\author{M.B.~Oktay}
\affiliation{Physics Department, University of Utah, Salt Lake City, Utah, USA}
\author{J.N.~Simone}
\affiliation{Fermi National Accelerator Laboratory, Batavia, Illinois, USA}
\author{R.~Sugar}
\affiliation{Department of Physics, University of California, 
Santa Barbara, California, USA}
\author{D.~Toussaint}
\affiliation{Department of Physics, University of Arizona, 
Tucson, Arizona, USA}
\author{R.S.~Van~de~Water}
\affiliation{Physics Department, Brookhaven National Laboratory, 
Upton, New York, USA}
\collaboration{Fermilab Lattice and MILC Collaborations}
\noaffiliation

\date{August 26, 2009}

\begin{abstract}
Comparisons of lattice-QCD calculations of semileptonic form factors 
with experimental measurements often display two sets of points, 
one each for lattice QCD and experiment.
Here we propose to display the output of a lattice-QCD analysis as a 
curve and error band.
This is justified, because lattice-QCD results rely in part on fitting, 
both for the chiral extrapolation and to extend lattice-QCD data over 
the full physically allowed kinematic domain.
To display an error band, correlations in the fit parameters must be 
taken into account.
For the statistical error, the correlation comes from the fit.
To illustrate how to address correlations in the systematic errors, 
we use the Be\'cirevi\'c-Kaidalov parametrization of the $D\to\pi l\nu$ 
and $D\to Kl\nu$ form factors, and an analyticity-based fit for the 
$B\to\pi l\nu$ form factor~$f_+$. 
\end{abstract}

\maketitle

The past several years have witnessed considerable improvement in our 
understanding of semileptonic decays of $D$ and $B$ mesons.
Measurements have advanced in accuracy from 6--20\% on the
normalization~\cite{Ablikim:2004ej,Huang:2004fra,Link:2004um} and 
$\sim10\%$ on the shape~\cite{Link:2004dh} to 
$\sim1\%$ on both~\cite{Besson:2009uv}.
Meanwhile, \emph{ab initio} calculations in QCD with lattice gauge 
theory have become realistic~\cite{Aubin:2004ej,Okamoto:2004xg,%
Dalgic:2006dt,Bailey:2008wp}, now incorporating the effects of sea 
quarks that were omitted in earlier work~\cite{Bowler:1999xn,%
Abada:2000ty,ElKhadra:2001rv,Aoki:2001rd,Shigemitsu:2002wh}.
In this article, we discuss how to present both together, so that 
the agreement (or, in principle, lack thereof) is easy to assess.

We~focus on reactions mediated by electroweak vector currents, leading to 
pseudoscalar mesons, $\pi$ or~$K$, in the final state.
At the quark level, a heavy quark $h$ decays into a daughter quark~$d$ 
(not necessarily the down quark), with a spectator antiquark~$\bar{q}$.
Writing the decay $H\to Pl\nu$, the form factors are defined by
\begin{equation}
    \langle P|V^\mu|H\rangle = 
        f_+(q^2) (p_H+p_P-\Delta)^\mu + f_0(q^2) \Delta^\mu  
    \label{eq:ff}
\end{equation}
where $q=p_H-p_P$ is the 4-momentum of the lepton system,
and $\Delta^\mu=(p_H+p_P)\cdot q\,q^\mu/q^2=(m_H^2-m_P^2)q^\mu/q^2$.
Equation~(\ref{eq:ff}) is general, applying to $K\to\pi l\nu$ as well 
as to $D$ and $B$ decays.
For lattice QCD, it is more convenient to express the transition 
matrix element as
\begin{equation}
    \langle P|V^\mu|H\rangle = \sqrt{2m_H} 
        \left[v^\mu f_\parallel(E) + p^\mu_\perp f_\perp(E) \right],
    \label{eq:ff-vp}
\end{equation}
where $v=p_H/m_H$, and $p_\perp=p_P-Ev$ and $E={v\cdot p_P}$ denote the 
3-momentum and energy of the final-state meson in the rest frame of the 
initial state.
The energy~$E$ is related to $q^2$ via
\begin{equation}
    q^2 = m_H^2 + m_P^2 - 2m_HE.
    \label{eq:q2E}
\end{equation}
Neglecting the lepton mass, $0\le q^2\le q^2_{\rm max}=(m_H-m_P)^2$ is 
kinematically allowed in the semileptonic decay.

The form factors~$f_+(q^2)$ and~$f_0(q^2)$ are related to~$f_\parallel(E)$ 
and~$f_\perp(E)$ by 
\begin{eqnarray}
    f_+(q^2) & = & (2m_H)^{-1/2}
        \left[f_\parallel(E) + (m_H-E)f_\perp(E) \right],
    \label{eq:f+} \\
    f_0(q^2) & = & \frac{\sqrt{2m_H}}{m_H^2-m_P^2}
        \left[(m_H-E) f_\parallel(E) - p_\perp^2 f_\perp(E) \right],
    \label{eq:f0} \hspace*{1.5em}
\end{eqnarray}
with Eq.~(\ref{eq:q2E}) understood.
Equations~(\ref{eq:f+}) and~(\ref{eq:f0}) imply $f_+(0)=f_0(0)$, as
required in Eq.~(\ref{eq:ff}).

Two aspects of lattice-QCD calculations are important here.
First (as in all lattice-QCD calculations), it is computationally 
demanding to have a spectator quark with mass as small as those of the 
up and down quarks; for $P=\pi$ the same applies to the daughter quark.
In recent unquenched calculations, the mass of the $\bar{q}q$ 
pseudoscalar $P_{\bar{q}q}$ lies in the range 
$0.1m_K^2\lesssim m_{P_{\bar{q}q}}^2 \lesssim m_K^2$.
Second (of special importance in semileptonic decays), the calculations 
take place in a finite spatial volume, so the 3-momentum takes discrete 
values.
In typical cases the box-size $L\approx2.5~\textrm{fm}$, so the smallest 
nonzero momentum $\bm{p}_{(1,0,0)}=2\pi(1,0,0)/L$ satisfies 
$|\bm{p}_{(1,0,0)}|\approx500~\textrm{MeV}$.

After generating numerical data at several values of 
$(E,m_{P_{\bar{q}q}}^2)$, the next step for lattice-QCD calculations is 
to carry out a chiral extrapolation, $m_{P_{\bar{q}q}}^2\to m_\pi^2$, of 
the data for $f_\perp$ and $f_\parallel$~\cite{Becirevic:2002sc,Aubin:2007mc}.
The chiral extrapolation must reflect the fact that the form factors are 
analytic in~$E=\sqrt{\bm{p}^2+m_P^2}$, not $\bm{p}$~\cite{Lellouch:1995yv}.
Note also that $f_\perp$ and $f_+$ can be computed only with 
$\bm{p}\neq\bm{0}$, hence $E>m_P$ or, equivalently, $q^2<q^2_{\rm max}$.
The statistical and discretization uncertainties in 
$f_+(q^2,m_{P_{\bar{q}q}}^2)$ start out smallest at $q_{(1,0,0)}^2$, 
corresponding to $\bm{p}_{(1,0,0)}$. 
A~sensible chiral extrapolation will propagate this feature to
$f_+(q^2,m_\pi^2)$.
Similarly, the statistical and discretization uncertainties in 
$f_0(q^2,m_\pi^2)$ are smallest near~$q^2_{\rm max}$.

When $|\bm{p}|a$ becomes too large, discretization effects grow out of 
control.
Therefore, the kinematic domain of lattice-QCD calculations is limited 
to, these days, $|\bm{p}|\lesssim 1~\textrm{GeV}$, with a corresponding 
upper limit on $E$ and lower limit on~$q^2$.
To extend the form factor over the full physical kinematic domain, 
a parametrization of the $q^2$ dependence is needed.

One choice is the Be\'cirevi\'c-Kaidalov (BK) 
ansatz~\cite{Becirevic:1999kt}
\begin{eqnarray}
    f_+(q^2) & = & \frac{F}{(1-\tilde{q}^2)(1-\alpha\tilde{q}^2)},
    \label{eq:BK+} \\
    f_0(q^2) & = & \frac{F}{1-\tilde{q}^2/\beta},
    \label{eq:BK0}
\end{eqnarray}
where $\tilde{q}^2=q^2/m_{H^*}^2$ ($H^*$ is the vector meson of 
flavor $h\bar{d}$), and $F$, $\alpha$, and $\beta$ are free parameters 
to be fitted.
A~key feature of Eq.~(\ref{eq:BK+}) is the built-in pole at 
$q^2=m_{H^*}^2$, or $E=-(m_{H^*}^2-m_H^2-m_P^2)/2m_H<0$, an indisputable 
feature of the physical~$f_+$. 
Further singularities at higher negative energy are modeled by the 
BK~parameters $\alpha$ and $\beta$.
A similar possibility is the Ball-Zwicky (BZ) 
ansatz~\cite{Ball:2004ye,Dalgic:2006dt}, 
which has one more parameter for $f_+$ than~BK.
A shortcoming of these parametrizations is that comparisons of 
lattice-QCD and experimental slope parameters can be 
misleading~\cite{Hill:2005ju,Becher:2005bg}, because 
lattice-QCD slopes are determined near $q^2=q^2_{\rm max}$, whereas
experimental slopes are determined near $q^2=0$.

Another approach based on analyticity and unitarity is to write the form 
factors as
\begin{eqnarray}
    f_+(q^2) & = & \frac{1}{(1-\tilde{q}^2)\phi_+(q^2)}\sum_{k=0}^Na_kz^k,
    \label{eq:z+} \\
    f_0(q^2) & = & \frac{1}{\phi_0(q^2)}\sum_{k=0}^Nb_kz^k,
    \label{eq:z0}
\end{eqnarray}
where $\phi_{+,0}$ are arbitrary, but suitable, functions, and the series 
coefficients are fit parameters.
The variable
\begin{equation}
    z = \frac{\sqrt{1-q^2/t_+}-\sqrt{1-t_0/t_+}}%
        {\sqrt{1-q^2/t_+}+\sqrt{1-t_0/t_+}},
\end{equation}
where $t_+=(m_H+m_P)^2$ and $t_0$ can be chosen to make $|z|$ small 
for all kinematically allowed $q^2$.
Like BK and BZ, Eq.~(\ref{eq:z+}) builds the $H^*$ pole into $f_+$,
but this approach is model independent because 
unitarity~\cite{Bourrely:1980gp,Boyd:1994tt,Lellouch:1995yv}
and heavy-quark physics~\cite{Becher:2005bg} impose bounds on 
$\sum_k|a_k|^2$, $\sum_k|b_k|^2$, and because kinematics set $|z|<1$.
Consequently, the series can be truncated safely, once additional terms 
are negligible compared to other uncertainties in the analysis.

In all approaches the output of an analysis of lattice-QCD form factors 
is a fit, usually a two-stage fit of chiral extrapolation followed by 
$q^2$~parametrization.
Clearly, the final fit describes a curve, and the error matrix of the fit 
parameters describes an error band.
Nevertheless, lattice-QCD results usually have been plotted as a set of 
points with error bars at fiducial values of $q^2$ (or $E$).
These points evoke the underlying discrete nature of the 3-momentum 
$\bm{p}$ but, in general, the chosen values of $q^2$ (or $E$) have 
nothing to do with the original discrete values of~$\bm{p}$.
A plot with a curve plus error band exhibits the same information, while 
giving a visually superior sense of the correlations between points on 
the curve.

The experimental measurements of $f_+(q^2)$ come from counting events in 
bins of $q^2$ and removing coupling and kinematic factors.
The analysis inevitably entails some fitting, to correct for acceptance, 
etc., but the postfit bins of $q^2$ faithfully mirror the input to such 
fits.

If one would like to compare the calculations with the measurements, it 
is appealing to represent one as a curve with error band, and the other 
as points with error bars.
Bearing the foregoing remarks in mind, it seems natural to draw the 
curve for lattice-QCD calculations.
A~few years ago, we prepared illustrative plots for $D\to Kl\nu$ with the 
Fermilab-MILC~\cite{Aubin:2004ej,Okamoto:2004xg} lattice-QCD 
calculations and FOCUS~\cite{Link:2004dh} and 
Belle~\cite{Widhalm:2006wz} measurements.
The intent was pedagogical, and we showed the plots at seminars and 
conferences~\cite{Kronfeld:2005fy}.

Unfortunately, the error band in that effort was impressionistic, 
not rigorous.
With the prospect of yet-more-precise results based on CLEO-$c$'s full 
accumulation of 818~pb$^{-1}$~\cite{Besson:2009uv}, we now present
a version that treats the error band as rigorously as possible.
We also prepare plots for $D\to\pi l\nu$ and $B\to\pi l\nu$.

As before we shall base the plots for $D$ decays on 
Ref.~\cite{Aubin:2004ej}.
The final result of this analysis consists of the BK parameters 
$(F,\alpha,\beta)$ and the $3\times3$ error matrix.
The full statistical error matrix is contained in a detailed, 
unpublished description of a BK-based analysis of $B\to\pi l\nu$ form 
factors~\cite{Okamoto:2005}.
The best fit, statistical errors, and systematic errors are tabulated 
in Table~\ref{tbl:fit}.
The statistical correlation matrices
$\rho_{ij}=\sigma^2_{ij}/(\sigma^2_{ii}\sigma^2_{jj})^{1/2}$ 
are tabulated in Table~\ref{tbl:corrl}.
The correlations among systematic errors are discussed below.

\begin{figure}[p]
    \centering
    \includegraphics[width=0.48\textwidth]{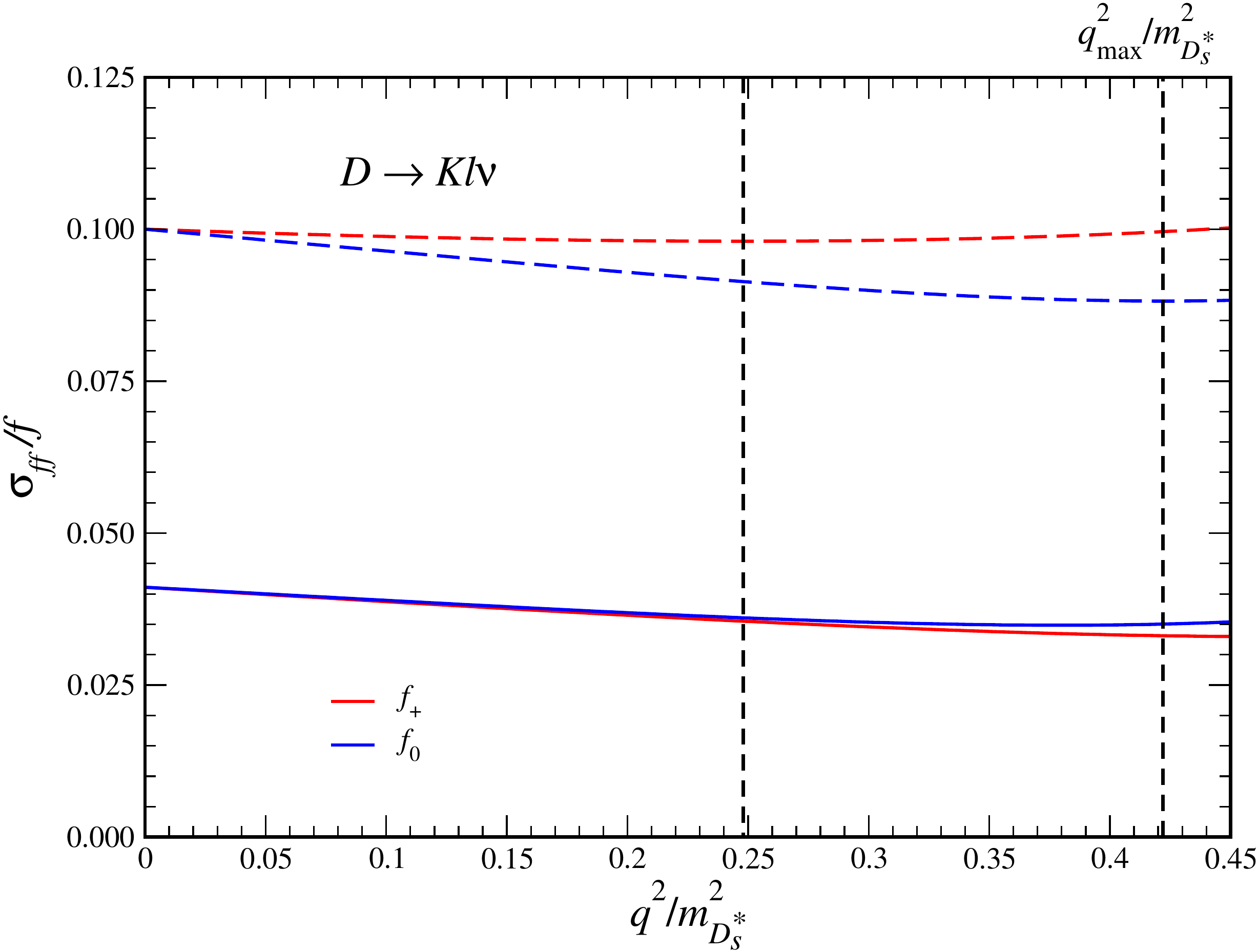} \\
    \includegraphics[width=0.48\textwidth]{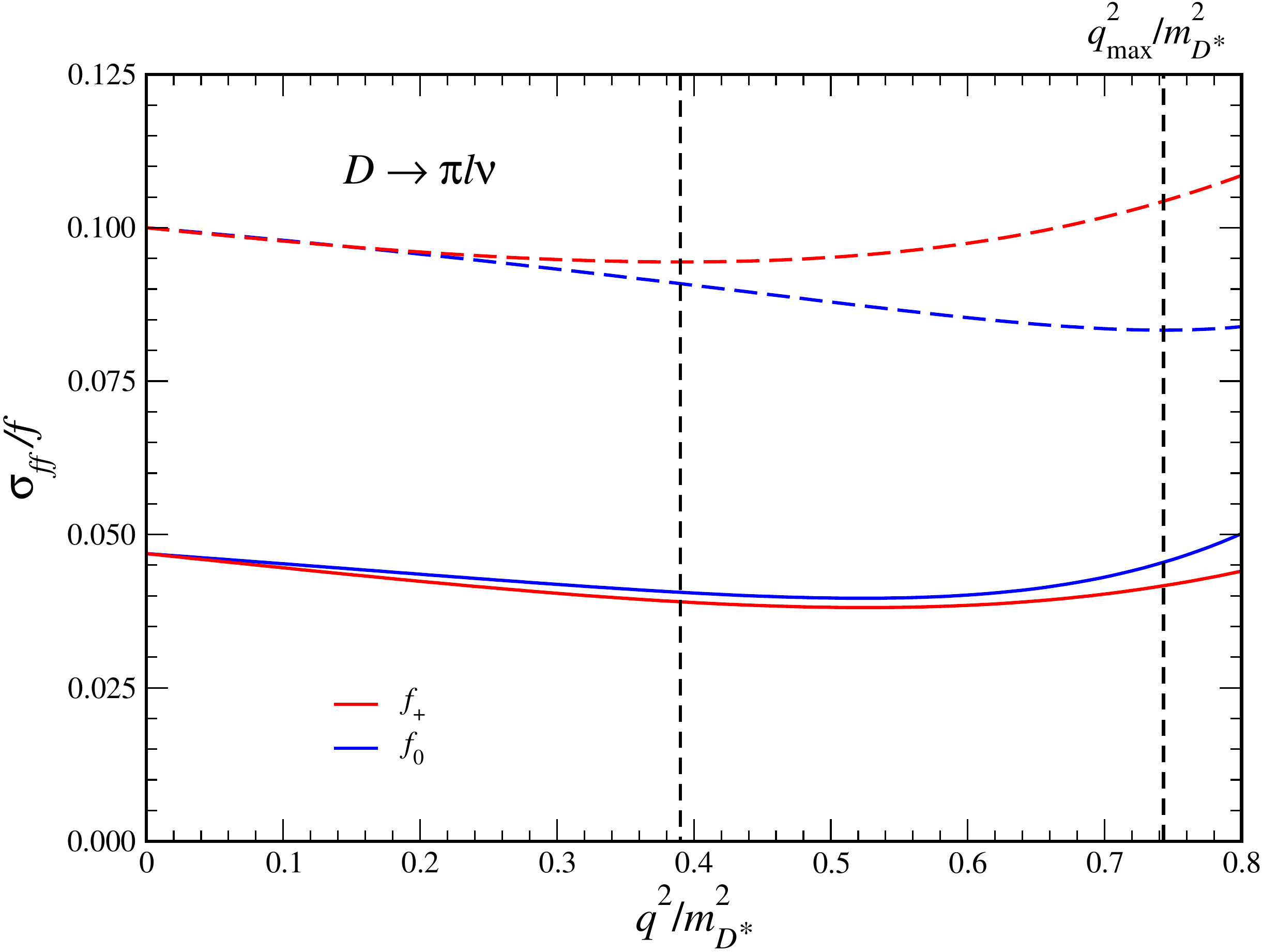}
    \caption{Relative errors vs~$q^2$.
        Solid (dashed) curves show the fitted statistical (estimated 
        systematic) error for $f_+$ (red curves) and $f_0$ (blue curves).
        Vertical lines show $q^2_{(1,0,0)}$ and $q^2_{\rm max}$.}
    \label{fig:sigma}
\end{figure}

Propagating (correlated) fluctuations in $F$, $\alpha$, and $\beta$ to 
the form factors, one finds relative squared-errors
\begin{eqnarray}
    \hspace*{-2.2pt}
    \frac{\sigma^2_{++}}{f_+^2} &  = &
    \frac{\sigma^2_{FF}}{F^2} + 
    2\frac{\sigma^2_{F\alpha}}{F}\frac{\tilde{q}^2}{1-\alpha\tilde{q}^{2}} +
    \sigma^2_{\alpha\alpha}
        \left(\frac{\tilde{q}^2}{1-\alpha\tilde{q}^{2}}\right)^2\!\!,
    \label{eq:f+error} \hspace*{0.7em} \\
    \frac{\sigma^2_{00}}{f_0^2} &  = &
    \frac{\sigma^2_{FF}}{F^2} - 
    2\frac{\sigma^2_{F\beta}}{F\beta} \frac{\tilde{q}^2}{\beta-\tilde{q}^{2}} +
    \frac{\sigma^2_{\beta\beta}}{\beta^2}
        \left(\frac{\tilde{q}^2}{\beta-\tilde{q}^{2}}\right)^2.
    \label{eq:f0error}
\end{eqnarray}
These errors are plotted as a function of $q^2$ in Fig.~\ref{fig:sigma} 
as solid curves.
\begin{table}[p]
    \centering
    \caption{Best-fit values of BK parameters with statistical and 
    systematic errors, successively, in
    parentheses~\cite{Aubin:2004ej,Okamoto:2004xg,Okamoto:2005}.}
    \label{tbl:fit}
    \begin{tabular}{clll}
        \hline\hline
              Decay    &  \quad$F$  & \quad$\alpha$ & \quad$\beta$ \\
        \hline
        $D\to K  l\nu$ & 0.73(3)(7) & 0.50(4)(7)    & 1.31(7)(13) \\
        $D\to\pi l\nu$ & 0.64(3)(6) & 0.44(4)(7)    & 1.41(6)(7) \\ 
        \hline\hline
\end{tabular}
\end{table}
\begin{table}[p]
    \centering
    \caption{Statistical error correlation matrices
        $\rho_{ij}=\sigma^2_{ij}/(\sigma^2_{ii}\sigma^2_{jj})^{1/2}$ 
        of the BK parameters~\cite{Okamoto:2005}.}
    \label{tbl:corrl}
    \begin{tabular}{crrr}
    \hline\hline
        $D\to K$ & \multicolumn{1}{c}{$F$} & 
            \multicolumn{1}{c}{$\alpha$} & \multicolumn{1}{c}{$\beta$} \\
           $F$   &    1.000 & $-$0.597 &    0.530 \\
        $\alpha$ & $-$0.597 &    1.000 & $-$0.316 \\
        $\beta$  &    0.530 & $-$0.316 &    1.000 \\   
    \hline
        $D\to\pi$ & \multicolumn{1}{c}{$F$} & 
            \multicolumn{1}{c}{$\alpha$} & \multicolumn{1}{c}{$\beta$} \\
           $F$   &    1.000 & $-$0.583 &    0.535 \\
        $\alpha$ & $-$0.583 &    1.000 & $-$0.312 \\ 
        $\beta$  &    0.535 & $-$0.312 &    1.000 \\
    \hline\hline
    \end{tabular}
\end{table}
\begin{figure}[bp]
    \centering
    \includegraphics[width=0.48\textwidth]{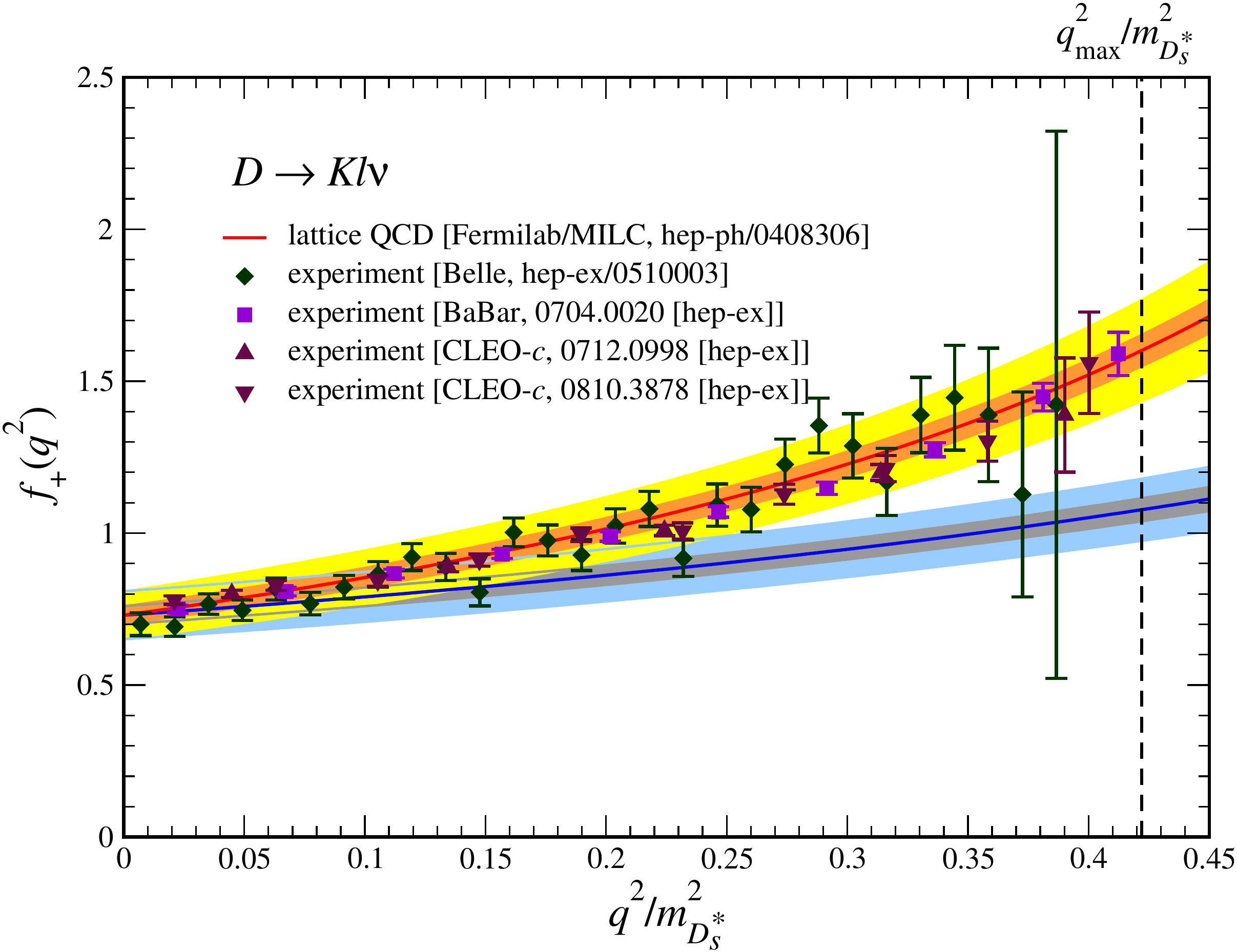} \\
    \includegraphics[width=0.48\textwidth]{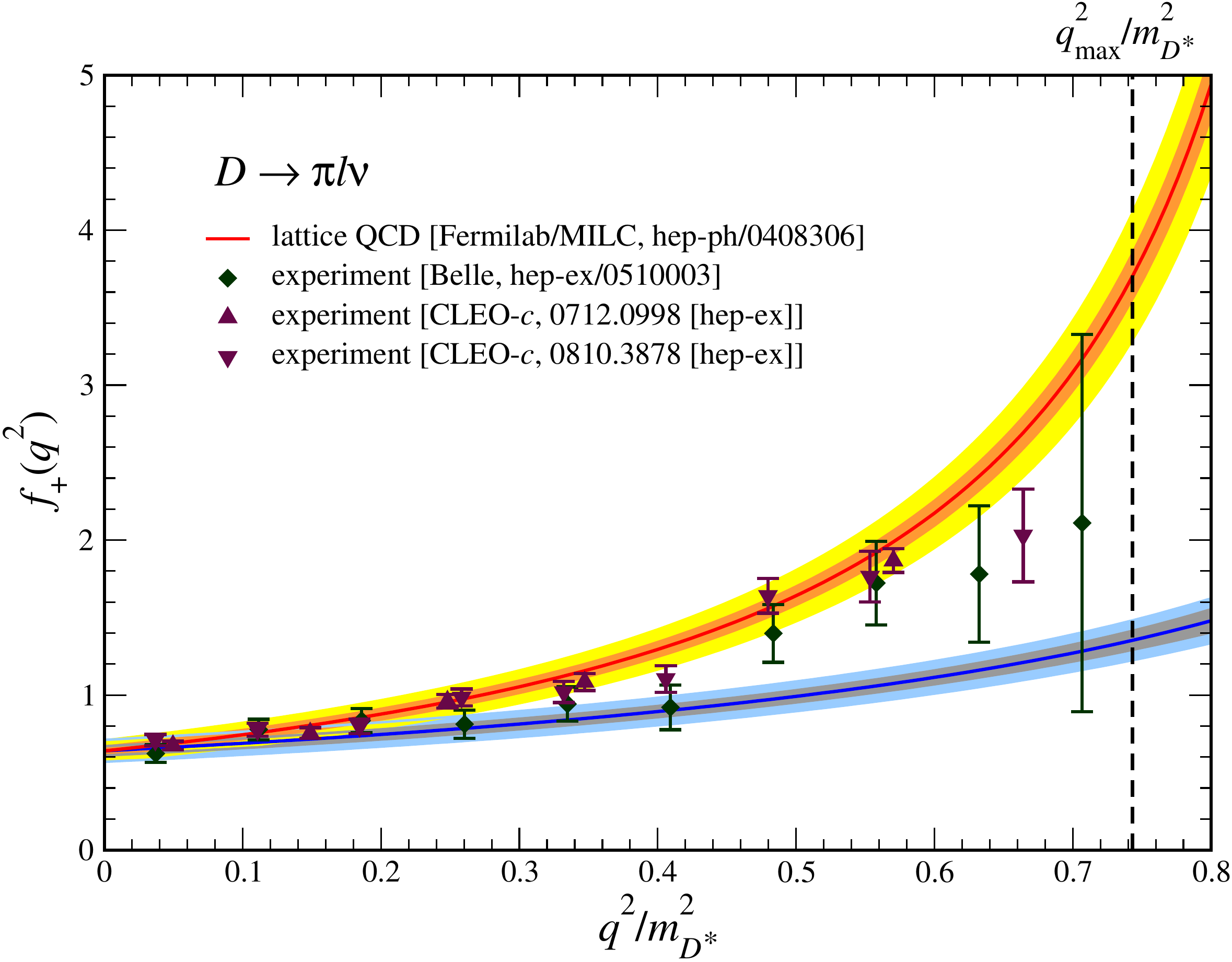}
    \caption{Form factors $f_+$ (and $f_0$) for semileptonic $D$ decays,
        from lattice QCD~\cite{Aubin:2004ej,Okamoto:2005},
        expressed as a red (blue) curve with an orange (gray)
        statistical error band and a yellow (light blue) combined 
        error band.
        Error bands take correlations into account.
        Measurements of $f_+$ are from 
        Belle (green diamonds)~\cite{Widhalm:2006wz},
        BaBar (magenta squares)~\cite{Aubert:2007wg}, and
        CLEO-$c$ (maroon triangles)~\cite{:2007se,:2008yi}.
        The vertical line shows $q^2_{\rm max}$.}
    \label{fig:ff}
\end{figure}
The relative statistical errors are smallest for $q^2$ such that
\begin{eqnarray}
    \sigma^2_{F\alpha} & = & - F\sigma^2_{\alpha\alpha}
        \tilde{q}^2/\left(1-\alpha\tilde{q}^{2}\right),
    \label{eq:f+errorMin} \hspace*{1em} \\
    \sigma^2_{F\beta} & = & F\sigma^2_{\beta\beta}
        \tilde{q}^2/\beta\left(\beta-\tilde{q}^{2}\right).
    \label{eq:f0errorMin}
\end{eqnarray}
It is illustrative to take $\sigma^2_{F\alpha}$ and $\sigma^2_{F\beta}$ 
from Tables~\ref{tbl:fit} and~\ref{tbl:corrl} and solve 
Eqs.~(\ref{eq:f+errorMin}) and~(\ref{eq:f0errorMin}) for $\tilde{q}^2$.
We~call these values $\tilde{q}^2_\alpha$ and $\tilde{q}^2_\beta$ and 
tabulate them, as well as $\tilde{q}_{(1,0,0)}^2$ and 
$\tilde{q}_{\rm max}^2$, in Table~\ref{tbl:stuff}.
\begin{table*}[tp]
    \centering
    \caption{Useful quantities for generating and assessing 
    Figs.~\ref{fig:sigma}, \ref{fig:ff}, and \ref{fig:Bpi}.}
    \label{tbl:stuff}
    \begin{tabular*}{0.8\textwidth}{@{\extracolsep{\fill}}cccccccccc}
        \hline\hline
        Decay & $H^*$ & $m_{H^*}$ & $E_{(1,0,0)}$ & 
        $q_{(1,0,0)}^2$ & $q_{\rm max}^2$ & 
        $\tilde{q}_{(1,0,0)}^2$ & $\tilde{q}_{\rm max}^2$ &
        $\tilde{q}_\alpha^2$ & $\tilde{q}_\beta^2$ \\
         &  & (MeV) & (MeV) & (GeV$^2$) & (GeV$^2$) & & & & \\
        \hline
        $D\to Kl\nu$ & $D_s^*$ & 2112 & 704 & 1.10 & 1.88 & 0.25 & 0.42 &
        0.47 & 0.38 \\
        $D\to\pi l\nu$ & $D^*$ & 2008 & 518 & 1.57 & 3.00 & 0.39 & 0.74 & 
        0.53 & 0.52 \\
        $B\to\pi l\nu$ & $B^*$ & 5325 & 518 & 22.4 & 26.4 & 0.79 & 0.93 & 
        --- & --- \\
        \hline\hline
    \end{tabular*}
\end{table*}
As one can see from Fig.~\ref{fig:sigma} and Table~\ref{tbl:stuff}, the 
statistical error is smallest between $q^2_{(1,0,0)}$ and~$q^2_{\rm max}$, 
as expected.
One may view this outcome as a check on the fitting procedures.

One can reverse this strategy to determine the correlation between the 
systematic errors of $F$ and the slope parameters.
In the error budget of Ref.~\cite{Aubin:2004ej} the largest systematic 
effect comes from discretization errors.
These should be smallest around $q_{(1,0,0)}^2$ and $q_{\rm max}^2$ for 
$f_+$ and $f_0$, respectively, because those correspond to the smallest 
$|\bm{p}|$ yielding the respective matrix elements.
This yields 
\begin{eqnarray}
    \rho_{F\alpha}^{\rm syst} & = & -0.198~(D\to K),
        \quad -0.329~(D\to\pi), \\
    \rho_{F\beta}^{\rm syst}  & = & +0.471~(D\to K),
        \quad +0.533~(D\to\pi), 
\end{eqnarray}
and the dashed curves in Fig.~\ref{fig:sigma}.

It is customary to combine statistical and systematic uncertainties by 
adding the two $\sigma^2$ (matrices).
Carrying out this procedure leads to the curves and bands in
Fig.~\ref{fig:ff}.
The error bands seem to contradict the conventional wisdom that the 
lattice-QCD uncertainties are smallest near $q^2_{\rm max}$.
This is not entirely the case for the \emph{relative} error, as seen in 
Fig.~\ref{fig:sigma}.
As~$q^2$ increases, the relative errors decrease until hitting a minimum
somewhere between $q^2_{(1,0,0)}$ and $q^2_{\rm max}$, as is reasonable.
The form factors rise faster than the relative errors drop, leading to 
the increasing absolute error seen in Fig.~\ref{fig:ff}.
These features are not an artifact of the BK parametrization, as we 
shall see below with the $B\to\pi l\nu$ form factor~$f_+$.

Figure~\ref{fig:ff} is the first main result of this article.
It shows the form factors $f_+$ and $f_0$ for $D\to Kl\nu$ and 
$D\to\pi l\nu$.
The lattice-QCD results are shown as curves (red for $f_+$, blue for 
$f_0$) with two errors bands, one statistical (orange for $f_+$, gray 
for $f_0$), the other systematic and statistical combined (yellow for 
$f_+$, light blue for $f_0$).
Experimental measurements for $f_+$
\cite{Widhalm:2006wz,Aubert:2007wg,:2007se,:2008yi}
are overlaid as points with error bars.
It may require careful scrutiny to see which experiment is which, 
but a glance reveals how well the points and curves agree.
The agreement is good for $D\to\pi l\nu$ and very good for $D\to Kl\nu$.

For the $z$ expansion the propagation of errors is even simpler.
Focusing on $f_+$, one has from Eq.~(\ref{eq:z+})
\begin{equation}
    \frac{\sigma^2_{++}}{f_+^2} = 
        \frac{\sum_{k,l=0}^N\sigma^2_{kl}z^{k+l}}%
        {\left[\sum_{k=0}^Na_kz^k\right]^2},
\end{equation}
where the indices on $\sigma^2$ correspond to those on the series 
coefficients.
The coefficients and error matrix for the $N=3$ fit ($t_0=0.65q^2_{\rm 
max}$) are tabulated in Table~\ref{tbl:fitz}.
\begin{table}[tbp]
    \centering
    \caption{Best-fit values $a_k$ and correlation matrix $\rho_{kl}$
    of the 3-term $z$ expansion of $f_+$ for $B\to\pi l\nu$, with 
    statistical and systematic errors combined~\cite{Bailey:2008wp}.}
    \label{tbl:fitz}
    \begin{tabular}{cccc}
        \hline\hline
            Fit:\quad & 0.0216(27) & $-0.0378(191)$ & $-0.113(27)$ \\
        \hline
            $\rho$ & \multicolumn{1}{c}{$a_0$} & 
                \multicolumn{1}{c}{$a_1$} & \multicolumn{1}{c}{$a_2$} \\
              $a_0$  &    1.000 &    0.640 &    0.475 \\
              $a_1$  &    0.640 &    1.000 &    0.964 \\
              $a_2$  &    0.474 &    0.964 &    1.000 \\
    \hline\hline
    \end{tabular}
\end{table}
The $z$-series fit was carried out after assigning $q^2$-dependent 
systematic uncertainties, so Table~\ref{tbl:fitz} refers to the combined 
statistical and systematic errors of this analysis.

\begin{figure}[bp]
    \includegraphics[width=0.48\textwidth]{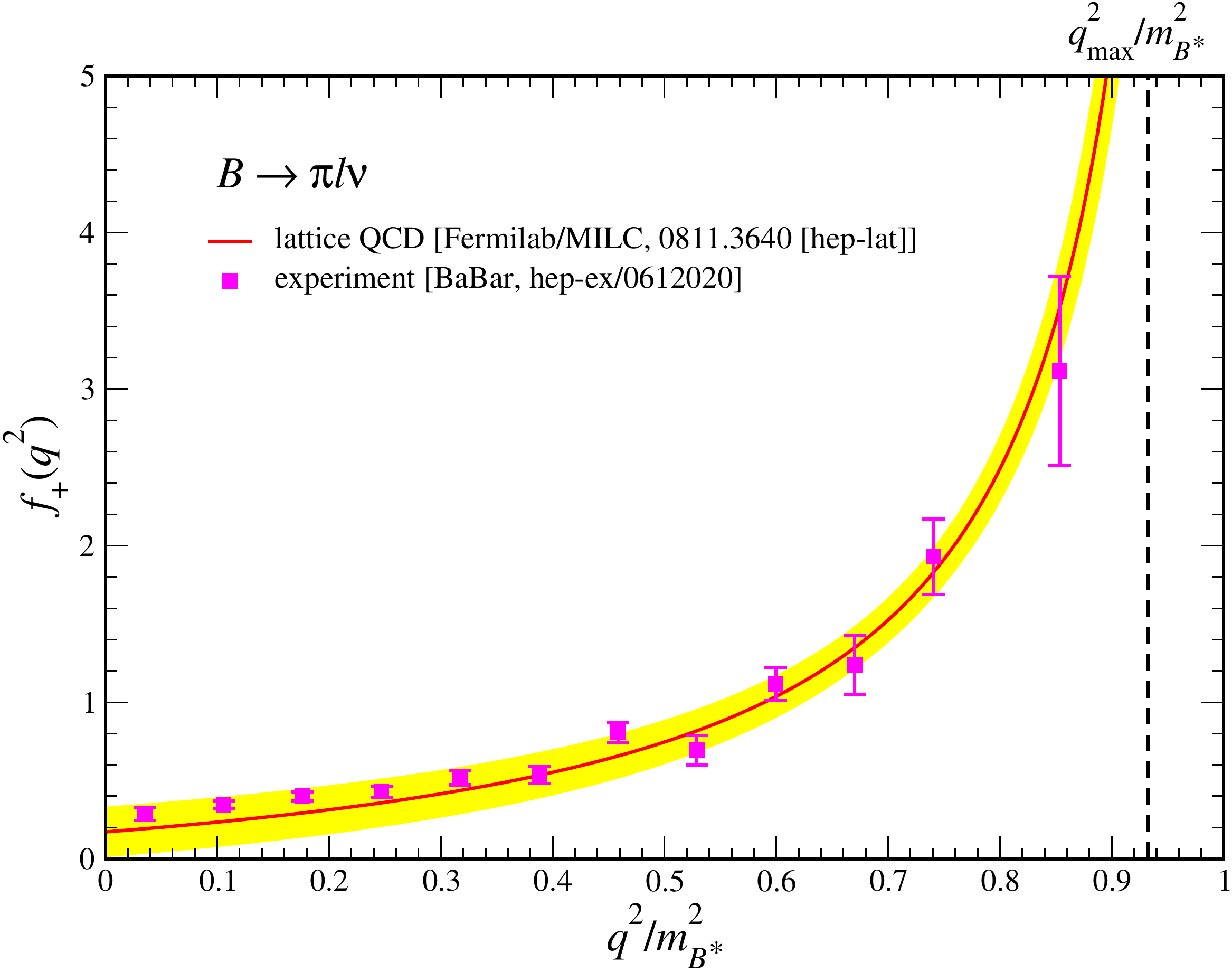}
    \caption{Form factor~$f_+$ for $B\to\pi l\nu$ expressed as a curve 
        (red) from the best fit with a total error band (yellow) from 
        taking correlations in the fit parameters into 
        account~\cite{Bailey:2008wp}, overlaid with measurements 
        of $|V_{ub}|f_+/(3.38\times10^{-3})$ from BaBar (magenta 
        squares)~\cite{Aubert:2006px}.
        The vertical line shows $q^2_{\rm max}$.}
    \label{fig:Bpi}
\end{figure}

This information, combined with the outer function~$\phi_+$ 
\cite{Bailey:2008wp}, is used to produce Fig.~\ref{fig:Bpi},
the second main result of this paper.
Now the curve and error band conform with preconceptions, for several 
reasons.
First, $q^2_{(1,0,0)}$ is close to $q^2_{\rm max}$, rather than in 
the middle of the kinematic range.
Second, the chiral extrapolation in Ref.~\cite{Bailey:2008wp} is less 
aggressive than that in Ref.~\cite{Aubin:2004ej}, leading to a larger 
but more realistic error at $q^2=0$.
The most striking aspect is that even though the absolute error in $f_+$ 
is increasing for $\tilde{q}^2\ge0.8$, the band remains narrow.
The band simply conveys the point-to-point correlations better.

This paper shows in detail how to compare semileptonic form factors from 
lattice-QCD and from experiments.
For illustration we use the BK parametrization for $D\to Kl\nu$ and 
$D\to\pi l\nu$, and the $z$ expansion for $B\to\pi l\nu$.
Clearly, the idea is more general.
For example, an interesting prospect relevant to semileptonic form 
factors is to inject 3-momenta smaller that $\bm{p}_{(1,0,0)}$ using 
``twisted'' boundary conditions~\cite{Bedaque:2004kc,%
Sachrajda:2004mi,Flynn:2005in}. 
That strategy should improve the accuracy of parameters in the chiral 
extrapolation and, hence, the BK, BZ, or $z$ fits.  
The output of any fit could still be exhibited as outlined here, 
although one should bear in mind that superior visualization of a 
fitting procedure does not repair any shortcomings of the fit itself.

\acknowledgments
We would like to thank Ian Shipsey for encouraging us to think  
carefully about the correlations in the systematic errors.
We would like to thank Laurenz Widhalm for providing the Belle data in 
numerical form~\cite{Widhalm:2006wz}, and Shipsey for the BaBar and CLEO 
$D$-decay data~\cite{Aubert:2007wg,:2007se,:2008yi}.
Computations for this work were carried out in part on facilities of
the USQCD Collaboration, which are funded by the Office of Science of
the United States Department of Energy.
This work was supported in part by the U.S. Department of Energy
under Grants No.~DE-FC02-06ER41446 (C.D., L.L., M.B.O.), 
No.~DE-FG02-91ER40661 (S.G.), No.~DE-FG02-91ER40677 (A.X.K., R.T.E., E.G.), 
No.~DE-FG02-91ER40628 (C.B., J.L.), and No.~DE-FG02-04ER41298 (D.T.); 
by the National Science Foundation under Grants No.~PHY-0555243, 
No.~PHY-0757333, No.~PHY-0703296 (C.D., L.L., M.B.O.), 
No.~PHY-0555235 (J.L.), and No.~PHY-0757035 (R.S.); 
and by Universities Research Associates (R.T.E., E.G.).
This manuscript has been coauthored by an employee of Brookhaven Science 
Associates, LLC, under Contract No. DE-AC02-98CH10886 with the U.S. 
Department of Energy.
Fermilab is operated by Fermi Research Alliance, LLC, under Contract 
No.~DE-AC02-07CH11359 with the U.S. Department of Energy.

\end{document}